\documentclass[11pt]{article}
\usepackage{amsmath,amssymb}
\usepackage{framed}
\usepackage{color}

\usepackage{graphicx}

\usepackage{amsmath,latexsym}

\begin{document}

\setcounter{page}{1}

\pagestyle{plain}

\begin{center}
\Large{\bf Viable Intermediate Inflation in the Mimetic DBI Model}\\
\small \vspace{1cm} {Narges
Rashidi}$^{a,b}$\footnote{n.rashidi@umz.ac.ir (Corresponding Author)} \quad and \quad {
Kourosh Nozari}$^{a,b}$\footnote{knozari@umz.ac.ir} \\
\vspace{0.5cm} $^{a}$ Department of Theoretical Physics, Faculty of Science,
University of Mazandaran,\\
P. O. Box 47416-95447, Babolsar, IRAN\\
\vspace{0.5cm} $^{b}$ ICRANet-Mazandaran,
University of Mazandaran,\\
P. O. Box 47416-95447, Babolsar, IRAN\\
\end{center}

\begin{abstract}
We study the intermediate inflation in the mimetic Dirac-Born-Infeld
model. By considering the scale factor as $a=a_{0}\exp(bt^{\beta})$,
we show that in some ranges of the intermediate parameters $b$ and
$\beta$, the model is free of the ghost and gradient instabilities.
We study the scalar spectral index, tensor spectral index, and
the tensor-to-scalar ratio in this model and compare the results with
Planck2018 TT, TE, EE+lowE+lensing +BAO +BK14 data at $68\%$ and
$95\%$ CL. In this regard, we find some constraints on the
intermediate parameters that lead to the observationally viable
values of the perturbation parameters. We also seek the non-gaussian
features of the primordial perturbations in the equilateral configuration. By performing the numerical analysis on
the nonlinearity parameter in this configuration, we show that the
amplitude of the non-gaussianity in the intermediate mimetic DBI
model is predicted to be in the range $-16.7<f^{equil}<-12.5$. We show that,
with $0<b\leq 10$ and $0.345<\beta<0.387$, we have an
instabilities-free intermediate mimetic DBI model that gives the
observationally viable perturbation and non-gaussianity parameters.
\end{abstract}
\newpage

\section{Introduction}

A new approach to General Relativity has been proposed by
Chamseddine and Mukhanov in 2003, called mimetic gravity. The
important property of their approach is that the conformal symmetry
is respected as an internal degree of freedom~\cite{Cham13}.
Chamseddine and Mukhanov in their interesting proposal, have written
the physical metric in terms of an auxiliary metric and a scalar
field. In fact, the metric is given by
\begin{eqnarray}\label{eq1}
g_{\mu\nu}= -\tilde{g}^{\varrho\sigma}\,\phi_{,\varrho}\,\phi_{,\sigma}
\,\tilde{g}_{\mu\nu}\,,
\end{eqnarray}
where the free non-dynamical scalar field $\phi$ encodes the
conformal mode of the gravity. Also, the definition (\ref{eq1})
shows that if we perform a Weyl transformation on the auxiliary
metric ($\tilde{g}_{\mu\nu}$), the physical metric ($g_{\mu\nu}$)
remains invariant. Using the equation (\ref{eq1}) gives the 
following constraint on the scalar field~\cite{Cham13}
\begin{equation}\label{eq2}
g^{\mu\nu}\phi_{,\mu}\phi_{,\nu}=-1\,.
\end{equation}
In the action of the mimetic gravity introduced in
Ref.~\cite{Cham13}, there is a contribution of the matter fields
coupled to $g_{\mu\nu}$ that leads to an extra term in the
Einstein's field equations. In the sense that the dependence of this
extra term to the scale factor is $a^{-3}$, it is considered as a
source of dark matter. The mimetic gravity scenario has been
explored in another mathematical approach, by adding the Lagrange
multipliers in the action of the theory~\cite{Gol13,Ham13}. In
Ref.~\cite{Bar13}, some ghost-free models of the mimetic gravity
have been discussed. The authors of Ref.~\cite{Cham14} have
considered a potential for the mimetic field in a Lagrange
multiplier approach, leading to some interesting results. In fact,
they have shown that if we take the appropriate potential terms, it
is possible to consider the mimetic field as inflaton, quintessence,
or phantom fields. The mimetic gravity has attracted a lot of
attention and the authors have extended it to the braneworld
scenario~\cite{Sad17}, non-minimal coupling
model~\cite{Myr15,Hos18}, $f(G)$ gravity~\cite{Ast15}, Horndeski
gravity~\cite{Co16,Arr15}, $f(R)$ theories~\cite{Noj14,Ast16,Noj16},
unimodular $f(R)$ gravity~\cite{Odi16} and Galileon
gravity~\cite{Hag14}. In studying the perturbations in mimetic
gravity models, it is necessary to have the ghost and gradient
instabilities-free models. In Ref.~\cite{Zhe17} it is shown that the
direct coupling between the curvature of the space-time and the
higher derivatives of the mimetic field can help to overcome such
instabilities in some ranges of the parameters space. See the papers
in
Refs.~\cite{Ij16,Cap14,Mir14,Mal14,Ram15,Lan15,Ram16,Arr16,Ac16,Hir17,Cai17,Tak17,Yos17}
for more works on the (in)stability issue.

On the other hand, to solve some problems of the standard model of
cosmology, the inflation paradigm has been introduced, where a
single canonical scalar field (inflaton) with a flat potential has
been considered. The flat potential causes the slow-roll of the
inflaton and enough exponential expansion of the early universe. In
this simple model, the dominant modes of the primordial
perturbations are predicted to be scale-invariant, adiabatic and
gaussian
~\cite{Star79,Star80,Mukh81,Gut81,Haw82,Star82,Gut82,Lin82,Alb82,Star83,Lin90,Lid00a,Lid97,Rio02,Lyt09,Mal03}.
However, there is a lot of attention to the extended inflation
models predicting the non-Gaussian distributed
perturbations~\cite{Mal03,Bar04,Che10,Fel11a,Fel11b,Noz15,Noz16,Noz17,Noz18b,Ras20}.

One interesting inflation model is the DBI (Dirac-Born-Infeld)
inflation~\cite{Sil04,Che07,Ali04,Che05,Noz13a,Ras18}. In the string
based DBI model, where the radial position of a $D3$ brane
characterizes the scalar field~\cite{Sil04,Ali04}, it is possible to
have large non-Gaussianity that many authors are interested
to~\cite{Ali04,Che05,Noz13a,Ras18,Li08,Li14,Naz16,Qiu16,Kum16,Cho15}.
Another string-based scalar field is the tachyon
field~\cite{Sen99,Sen02} which is an interesting scalar field in the
inflation models~\cite{Ras18,Noz13b,Ras21}. We have included the DBI and
tachyon field in the mimetic models in our previous works. In
Ref.~\cite{Noz19}, we have assumed that the scalar field of the DBI
model is a mimetic scalar field. By considering the Lagrange
multiplier approach and assuming the power-law inflation, we have
studied the inflation and perturbations in this model and shown that
this model in some ranges of the model's parameter space is free of
the ghost and gradient instabilities. We have also shown that, in
those ranges of the model's parameter space, the values of the
perturbations parameters are observationally viable. In fact, with
the Mimetic DBI (MDBI) model, there is no need to consider such
complicated higher-order terms in the action of mimetic gravity. In
Ref.~\cite{Ras20b}, we have considered the tachyon field in the
mimetic gravity setup and within the Lagrange multiplier approach.
We have adopted both power-law and intermediate scale factors and
studied the inflation in the tachyon mimetic model. We have shown
that, in both cases, the mimetic model is free of instabilities in
some ranges of the model's parameters. Also, those ranges of the
parameters that lead to instabilities free tachyon mimetic model,
give observationally viable perturbation parameters.

In the continuation of the previous works, in this paper, we study
other aspects of the MDBI model. Now, we consider the MDBI model
with an intermediate scale factor~\cite{Bar90,Bar93,Bar07} that
gives another type of potential. In this regard, we show that the
intermediate MDBI model also is an instabilities-free mimetic model
that is consistent with observational data. Therefore, with this new
work on the MDBI model, we prove that to have an instabilities-free
mimetic model there is no need to restrict ourselves  to one type of
potential (or scale factor). The MDBI model is a simple mimetic
model that gives a viable cosmological model with both power-law
and intermediate scale factor (corresponding to two types of
potential). This makes the MDBI model more favorable. Also, in this
paper, we study the non-gaussian features of the primordial
perturbations that is one of the important issues in the
inflationary models. The prediction of the intermediate MDBI model
for the sound speed of the perturbations and also the amplitude of
the perturbations in equilateral configuration is
the issue that hasn't been studied in the MDBI model previously.

This paper is organized as follows: In section 2, we review the MDBI
model in the Lagrange multiplier approach. In this section, we
present the perturbation parameters (the scalar spectral index, the
tensor spectral index, and the tensor-to-scalar ratio) that are
functions of the Lagrange multiplier. The amplitude of the
non-gaussianity in the equilateral configuration is
also presented in this section, which is related to the Lagrange
multiplier via the sound speed of the model. In section 3, we
introduce the intermediate MDBI model where the scale factor is
given by $a=a_{0}\exp(bt^{\beta})$. By this scale factor, we find
the Hubble parameter in terms of the intermediate parameters $\beta$
and $b$. Then, by obtaining the potential and the Lagrange
multiplier in terms of the Hubble parameter, we find the slow-roll
parameters in the intermediate MDBI model. Also in this section, we
study the model numerically and show that the intermediate MDBI
model in some ranges of its parameter space is free of the ghost and
gradient instabilities. In section 4, by performing numerical
analysis, we compare the results with Planck2018 observational data.
In this regard, we show that the intermediate MDBI model for some
values of $\beta$ and $b$, which lead to the instabilities-free
model, gives observationally viable values of the perturbation
parameters. We also present some predictions of the model on the
non-gaussian feature of the primordial perturbations. In section 5,
we present a summary of our work.

\section{Mimetic DBI Model}

The action of the DBI mimetic gravity, in the presence of the
Lagrange multiplier and a potential term, is given by
\begin{eqnarray}
\label{eq3} S=\int
d^{4}x\sqrt{-g}\Bigg[\frac{R}{2\kappa^{2}}-{\cal{F}}^{-1}(\phi)\sqrt{1+{\alpha}{\cal{F}}(\phi)\partial_{\nu}\phi\,\partial^{\nu}\phi}
+\lambda(g^{\mu\nu}\partial_{\mu}\phi\,\partial_{\nu}\phi+1)-V(\phi)
\Bigg],
\end{eqnarray}
where $R$ is the Ricci scalar, $V(\phi)$ presents the potential of
the scalar field, ${\cal{F}}^{-1}(\phi)$ is the inverse brane
tension (note that, the D3-brane passes a compact manifold which
geometry of its throat is related to the ${\cal{F}}^{-1}(\phi)$).
Also, $\kappa$ is the gravitational constant, defined as $\kappa^{2}=\frac{8\pi G}{c^{4}}$. The parameter $\alpha$ is a coupling parameter which is corresponding to the string theory parameter. Although we adopt a DBI-like Lagrangian in our work, the scalar field $\phi$ is not a DBI field. It is a mimetic field obeying the constraint (\ref{eq2}). This means that its dimension is $[T]=[M]^{-1}$, demonstrating the dimension of time (note that we work in natural units where the speed of light is $c=1$ and also $\hbar=1$). Since the action should be dimensionless, to have dimensional consistency, there should be a coupling parameter $\alpha$ with dimension $[M]^4=[T]^{-4}$ in the Lagrangian term, as shown in equation (\ref{eq3}). This parameter is corresponding to the string theory parameter presented in \cite{Sil04}. Also, $\lambda$ is a Lagrange multiplier by which we enter the
mimetic constraint (\ref{eq2}) in the action.

If we vary action (\ref{eq3}) with respect to the metric, we find
the following Einstein's field equations
($G_{\mu\nu}=\kappa^{2}T_{\mu\nu}$)
\begin{eqnarray}
\label{eq4}
G_{\mu\nu}=\kappa^{2}\Bigg[-g_{\mu\nu}{\cal{F}}^{-1}\sqrt{1+{\alpha}{\cal{F}}\,g^{\mu\nu}\,\partial_{\mu}\phi\,\partial_{\nu}\phi}-g_{\mu\nu}V
+g_{\mu\nu}\,\lambda\Big(g^{\mu\nu}\,\partial_{\mu}\phi\,\partial_{\nu}\phi+1\Big)-2\lambda\,\partial_{\mu}\phi\,\partial_{\nu}\phi
\nonumber\\+{\alpha}\partial_{\mu}\phi\,\partial_{\nu}\phi
\Big(1+{\cal{F}}\,g^{\mu\nu}\,\partial_{\mu}\phi\,\partial_{\nu}\phi\Big)^{-\frac{1}{2}}\Bigg]\,.
\end{eqnarray}
By using the flat FRW metric as the background,
\begin{equation}
\label{eq5} ds^{2}=-dt^{2}+a^{2}(t)\delta_{ij}dx^{i}dx^{j}\,,
\end{equation}
the field equations (\ref{eq4}) give the following Friedmann
equations in this model
\begin{eqnarray}
\label{eq6}
3H^{2}=\kappa^{2}\Bigg[\frac{{\cal{F}}^{-1}}{\sqrt{1-{\alpha}{\cal{F}}\dot{\phi}^{2}}}+V-\lambda\Big(1+\dot{\phi}^{2}\Big)\Bigg]\,,
\end{eqnarray}

\begin{eqnarray}
\label{eq7}
2\dot{H}+3H^{2}=\kappa^{2}\Bigg[{\cal{F}}^{-1}\,\sqrt{1-{\alpha}{\cal{F}}\dot{\phi}^{2}}+V+\lambda\Big(\dot{\phi}^{2}-1\Big)\Bigg]\,.
\end{eqnarray}
The equation of motion of the mimetic field $\phi$ in the MDBI model
is obtained by varying the action (\ref{eq3}) with respect to $\phi$
\begin{eqnarray}
\label{eq8}\frac{{\alpha}\ddot{\phi}}{(1-{\alpha}{\cal{F}}\dot{\phi}^{2})^{\frac{3}{2}}}+\frac{3H{\alpha}\dot{\phi}}{(1{\alpha}{\cal{F}}\dot{\phi}^{2})^{\frac{1}{2}}}
-2\lambda\Big(\ddot{\phi}+3H\dot{\phi}\Big)+V'
-\lambda'(1-\dot{\phi}^{2})=-\frac{{\cal{F}}'}{{\cal{F}}^{2}}
\Bigg[\frac{3{\alpha}{\cal{F}}\dot{\phi}^{2}-2}{2(1-{\alpha}{\cal{F}}\dot{\phi}^{2})^{\frac{1}{2}}}\Bigg]\,.
\end{eqnarray}
Also, by varying action (\ref{eq3}) with respect to $\lambda$ we
reach the constraint (\ref{eq2}).

In this paper, we are going to consider the MDBI setup as an
inflation model. The slow-roll parameters in the inflation model are
given by the following definitions
\begin{equation}
\label{eq9}\epsilon\equiv-\frac{\dot{H}}{H^{2}}\,,\quad
\eta\equiv\frac{1}{H}\frac{d \ln \epsilon}{dt}\,,\quad
s\equiv\frac{1}{H}\frac{d \ln c_{s}}{dt}\,,
\end{equation}
where $c_{s}$ is the sound speed of the perturbations, defined as
$c_{s}^{2}=\frac{P_{,X}}{\rho_{,X}}$ (the subscript ``$,X$" shows
derivative with respect to
$X=-\frac{1}{2}\partial_{\nu}\phi\,\partial^{\nu}\phi$). In our
model, the square of sound speed is given by
\begin{eqnarray}
\label{eq10}c_{s}^{2}=\frac{{\alpha}\left(1-{\alpha}{\cal{F}}
\dot{\phi}^{2}\right)^{-\frac{1}{2}}-2\lambda}{{\alpha}\left(1-{\alpha}{\cal{F}}\dot{\phi}^{2}\right)^{-\frac{3}{2}}-2\lambda}\,,
\end{eqnarray}
which should satisfy the constraint $0<c_{s}^{2}\leq c^2$ (note that
$c$ is the local speed of light). In the case of $c\equiv 1$, the
constraint becomes as $0<c_{s}^{2}\leq 1$~\cite{Eli07,Qui17}.

To seek the observational viability of the MDBI model, it is
important to compare the values of the perturbation parameters with
Planck2018 data~\cite{pl18a,pl18b}. The Planck collaboration has
considered that in the case of the statistical isotropy, the
two-point correlations of the CMB anisotropies are described by the
angular power spectra~\cite{kam97,zal30,sel97,Hu97,Hu98}. In fact,
the following expressions for the contributions from the scalar and
tensor perturbations in the CMB angular power spectra have been
used~\cite{pl15}
\begin{eqnarray}
\label{eq11} C_{l}^{ab,s}=\int_{0}^{\infty} \frac{dk}{k}
\Delta_{l,a}^{s} (k) \, \Delta_{l,b}^{s} (k)\, {\cal{A}}_{s} (k)\,,
\end{eqnarray}

\begin{eqnarray}
\label{eq12} C_{l}^{ab,T}=\int_{0}^{\infty} \frac{dk}{k}
\Delta_{l,a}^{T} (k) \, \Delta_{l,b}^{T} (k)\, {\cal{A}}_{T} (k)\,,
\end{eqnarray}
where $a,b=T,E,B$. Also, $l$ is the multipole moment number,
$\Delta_{l,A}^{s}$ and $\Delta_{l,A}^{T}$ are the transfer
functions\footnote[2]{These functions are computed by using the
Boltzmann codes such as CMBFAST~\cite{sel96} or CAMB~\cite{Lew00}.},
and ${\cal{A}}_{j}(k)$ ($j=d,T$) is the primordial power spectrum,
identified by the physics of the primordial universe~\cite{pl15}. In
one procedure to compare the inflationary parameters with data, the
Planck collaboration has expanded the scalar and tensor power
spectra in a model-independent form as~\cite{pl18b,pl15}
\begin{eqnarray}
\label{eq13} {\cal{A}}_{s}
(k)=A_{s}\left(\frac{k}{k_{*}}\right)^{n_{s}-1+\frac{1}{2}\frac{dn_{s}}{d\ln
k}\ln\big(\frac{k}{k_{*}}\big)+\frac{1}{6}\frac{d^{2}n_{s}}{d\ln
k^{2}}\ln\big(\frac{k}{k_{*}}\big)^{2}+...}\,,\nonumber\\
\end{eqnarray}
\begin{eqnarray}
\label{eq14} {\cal{A}}_{T}
(k)=A_{T}\left(\frac{k}{k_{*}}\right)^{n_{T}+\frac{1}{2}\frac{dn_{T}}{d\ln
k}\ln\big(\frac{k}{k_{*}}\big)+...}\,,
\end{eqnarray}
where, $A_{j}$ is introduced as the amplitude of the scalar (for
$j=s$) or tensor (for $j=T$) perturbations. Also,
$\frac{dn_{j}}{d\ln k}$ is the the running of the scalar (for $j=s$)
or tensor (for $j=T$) spectral index and $\frac{d^{2}n_{s}}{d\ln
k^{2}}$ is the running of the running of the scalar spectral index.
The ratio between the tensor and scalar amplitudes
\begin{eqnarray}
\label{eq15} r=\frac{{\cal{A}}_{T}(k_{*})}{{\cal{A}}_{s}(k_{*})}\,,
\end{eqnarray}
is an important perturbation parameter, named tensor-to-scalar
ratio.

To perform numerical analysis and compare the results with
Planck2018 data, we should obtain the perturbation parameters in the
MDBI model. The scale dependence of the scalar spectral index, at
the time of sound horizon exit of the physical scales ($c_{s}k=aH$),
is identified by
\begin{eqnarray}
\label{eq16} n_{s}-1=\frac{d \ln {\cal{A}}_{s}}{d\ln
k}\Bigg|_{c_{s}k=aH}\,.
\end{eqnarray}
Calculating the perturbation parameters at pivot scale $k=k_{*}$
causes that the running term doesn't appear in the definition
(\ref{eq16}). For the MDBI model, the amplitude of the scalar
spectral index is given by~\cite{Noz19}
\begin{equation}
\label{eq17}{\cal{A}}_{s}=\frac{H^{2}}{8\pi^{2}{\cal{W}}_{s}c_{s}^{3}}\,,
\end{equation}
where
\begin{equation}
\label{eq18} {\cal{W}}_{s}=\frac {\left(
{\cal{F}}\,{\alpha}\,\dot{\phi}^{2}\lambda\,\sqrt
{1-{\alpha}{\cal{F}}\dot{\phi}^{2}}-\lambda\,\sqrt
{1-{\alpha}{\cal{F}}\dot{\phi}^{2}}+{\alpha} \right) \dot{\phi}^{2}}{2{H}^{2}
\left( 1-{\alpha}{\cal{F}}\dot{\phi}^{2} \right) ^{3/2}}\,.
\end{equation}
Positive values of ${\cal{W}}_{s}$ makes the model free of the ghost
instability. Now, we find the scalar spectral index as follows
\begin{eqnarray}
\label{eq19} n_{s}=1-6\epsilon+2\eta-s\,,
\end{eqnarray}
which is expressed in terms of the slow-roll parameters.

Equation (\ref{eq14}) gives the following tensor spectral index
\begin{eqnarray}
\label{eq20} n_{T}=\frac{d \ln {\cal{A}}_{T}}{d\ln
k}\Bigg|_{k=aH}\,.
\end{eqnarray}
The parameter ${\cal{A}}_{T}$, the amplitude of the tensor
perturbations, is given by
\begin{eqnarray}
\label{eq21} {\cal{A}}_{T}=\frac{2\kappa^{2}H^{2}}{\pi^{2}}\,,
\end{eqnarray}
leading to
\begin{eqnarray}
\label{eq22} n_{T}=-2\epsilon\,.
\end{eqnarray}
We can find the tensor-to-scalar ratio from equations (\ref{eq15})
and (\ref{eq22}) as follows
\begin{eqnarray}
\label{eq23} r=16\,c_{s}\,\epsilon\,,
\end{eqnarray}
or
\begin{eqnarray}
\label{eq24} r=-8\,c_{s}\,n_{T}\,.
\end{eqnarray}
Equation (\ref{eq24}) is an important equation, named the
consistency relation, which in the simple single filed inflation
with a canonical scalar field simplifies to $r=16\epsilon$.

To get more information about the viability of an inflation model,
it is useful to study the non-gaussian feature of the primordial
perturbations. Although the two-point correlation characterizes the
gaussian perturbations, to get the additional statistical
information related to the non-gaussian distribution, we should
consider three and higher-order correlations. In the interaction
picture, the 3-point correlation for the spatial curvature
perturbation $\Psi$ is given by~\cite{Mal03,Fel11b,See05}
\begin{eqnarray}
\label{eq25} \langle
{\Psi}(\textbf{k}_{1})\,{\Psi}(\textbf{k}_{2})\,{\Psi}(\textbf{k}_{3})\rangle
=(2\pi)^{3}\delta^{3}(\textbf{k}_{1}+\textbf{k}_{2}+\textbf{k}_{3}){\cal{B}}_{\Psi}(\textbf{k}_{1},\textbf{k}_{2},\textbf{k}_{3})\,,
\end{eqnarray}
where
\begin{equation}
\label{eq26}
{\cal{B}}_{\Psi}(\textbf{k}_{1},\textbf{k}_{2},\textbf{k}_{3})=\frac{(2\pi)^{4}{\cal{A}}_{s}}{\prod_{i=1}^{3}
k_{i}^{3}}\,
{\cal{G}}_{\Psi}(\textbf{k}_{1},\textbf{k}_{2},\textbf{k}_{3})\,,
\end{equation}
and ${\cal{A}}_{s}$ is given by (\ref{eq17}). The parameter
${\cal{G}}_{\Psi}$ in equation (\ref{eq26}) is defined as
\begin{equation}
\label{eq27}
{\cal{G}}_{\Psi}=\frac{3}{4}\Bigg(1-\frac{1}{c_{s}^{2}}\Bigg){\cal{S}}_{1}+\frac{1}{4}\Bigg(1-\frac{1}{c_{s}^{2}}\Bigg){\cal{S}}_{2}
+\frac{3}{2}\Bigg(\frac{1}{c_{s}^{2}}-1\Bigg){\cal{S}}_{3}\,,
\end{equation}
where the shapes of the non-gaussianity are given by
\begin{equation}
\label{eq28}
{\cal{S}}_{1}=\frac{2}{K}\sum_{i>j}k_{i}^{2}\,k_{j}^{2}-\frac{1}{K^{2}}\sum_{i\neq
j}k_{i}^{2}\,k_{j}^{3}\,,
\end{equation}
\begin{equation}
\label{eq29}
{\cal{S}}_{2}=\frac{1}{2}\sum_{i}k_{i}^{3}+\frac{2}{K}\sum_{i>j}k_{i}^{2}\,k_{j}^{2}-\frac{1}{K^{2}}\sum_{i\neq
j}k_{i}^{2}\,k_{j}^{3}\,,
\end{equation}
\begin{equation}
\label{eq30}
{\cal{S}}_{3}=\frac{\left(k_{1}\,k_{2}\,k_{3}\right)^{2}}{K^{3}}\,,
\end{equation}
and
\begin{equation}
\label{eq31} K=k_{1}+k_{2}+k_{3}\,.
\end{equation}
Equation (\ref{eq26}) shows that the three-point correlator depends
on the three momenta $k_{1}$ , $k_{2}$ and $k_{3}$. Note that these
momenta should satisfy the translation and rotational invariance. By
defining the following dimensionless parameter, called
``nonlinearity parameter'',
\begin{equation}
\label{eq32}
f=\frac{10}{3}\frac{{\cal{G}}_{\Psi}}{\sum_{i=1}^{3}k_{i}^{3}}\,,
\end{equation}
we can measure the amplitude of the non-gaussianity. The nonlinearity parameter depends on the shape of the non-gaussianity. Different values of the momenta lead to different shapes and there is a maximal signal for each shape in a special configuration of the
three momenta.

However, as has been demonstrated in
Refs.~\cite{Che07,Bab04b,Fel13a,Bau12}, for the $k$-inflation and
higher-order derivative models, the maximal peak of the signal
occurs at the equilateral configuration, where $k_{1}=k_{2}=k_{3}$.
In this regard, here also, we obtain the parameter
${\cal{G}}_{\Psi}$ in this configuration and at the leading-order
as~\cite{Che07,Ali04,Che05}
\begin{equation}
\label{eq33}
{\cal{G}}_{\Psi}^{equil}=\frac{17}{72}k^3\left(1-\frac{1}{c_{s}^{2}}\right)\,,
\end{equation}
leading to the following nonlinearity parameter
\begin{equation}
\label{eq34}
f^{equil}=\frac{85}{324}\left(1-\frac{1}{c_{s}^{2}}\right)\,.
\end{equation}
This equation is very important to explore the non-gaussian feature
of the primordial perturbations in an inflation model and compare
with observational data.

In the following, we study the intermediate inflation in the MDBI
model and perform some numerical analysis on this model.

\section{Intermediate MDBI Model}
One of the interesting scenarios in inflation models is
intermediate inflation. In the intermediate inflation, the scale
factor evolves faster than the power-law inflation ($a=t^{p}$) but
slower than the standard de Sitter inflation
($a=\exp(Ht)$)~\cite{Bar90,Bar93,Bar07}. In fact, in the
intermediate inflation the evolution of the scale factor is given by
\begin{eqnarray}
\label{eq35}a=a_{0}\,\exp\left(b\,t^{\beta}\right)\,,
\end{eqnarray}
where b is a constant and $0<\beta<1$. This scale factor leads to
the following Hubble parameter
\begin{eqnarray}
\label{eq36}H(N)=N \left( {\frac {N}{b}} \right)
^{-{\frac{1}{\beta}}}\beta\,.
\end{eqnarray}
To obtain the main perturbation parameters in the intermediate
inflation, we should follow~\cite{Bam14,Odi15} and find the
potential in terms of the Hubble parameter and its derivatives. From
now on, we assume ${\cal{F}}^{-1}(\phi)=V(\phi)$.

Note that, in Ref.~\cite{Ali04}, it has been demonstrated that for
$AdS_5$ throat, the warp factor ${\cal{F}}^{-1}(\phi)$ is equal to
$\frac{\varsigma}{\phi^{4}}$. Also, in the case with $AdS_5 \times X$
geometry, the potential of a DBI field is quartic. Another
interesting case is clarified in Refs.~\cite{Tsuj13,Kin08}. In those
papers, the authors have shown that with ${\cal{F}}\sim e^{m\phi}$
and $V\sim e^{-m\phi}$ (with $m$ to be a constant), we can get the
Lagrangian of the DBI model. Following Refs.~\cite{Tsuj13,Kin08}, we
consider ${\cal{F}}^{-1}(\phi)=V(\phi)$ in our calculations. In
fact, it is always possible to have viable DBI inflation with these
choices of functions. However, in this paper, we construct the
potential (and therefore, ${\cal{F}}$) for intermediate inflation
from background equations. In this way, there is no need to choose
an arbitrary function for $V$ and ${\cal{F}}$.

Now, we introduce a new scalar field $\varphi$. This field is
identified by the number of e-folds $N$ and parameterizes the scalar
field $\phi$ as $\phi=\phi(\varphi)$. From these points and by using
equation (\ref{eq7}), we obtain the potential in the intermediate
inflation as follows
{
\begin{eqnarray}
\label{eq37} V=\frac {\Big[2H H' +3H^{2} \Big]^{2}}{{\kappa}^{2}
\Big[-{\kappa}^{2}\,{\alpha}+4H H'+6H^{2}
 \Big] }\,.
\end{eqnarray}}
Note that, in the above and forthcoming equations, a prime refers to
the derivative of the parameter with respect to $N$, and also we have
$H=H(N)$. From equations (\ref{eq6}) and (\ref{eq37}) we get
{
\begin{eqnarray}
\label{eq38} \lambda=\frac
{9\,{H}^{4}{\cal{X}}-3\,{H}^{2}{\cal{X}}\alpha{\kappa}^{2}-9\,{H}^{4}-
12\,{H}^{3}H'}{-2{\cal{M}}{\cal{X}}{\kappa} ^{2}} +\frac
{-4\,H^{2}H'^{2}{\cal{X}}-4\,H^{2}H'^{2}}{-2{\cal{M}}{\cal{X}}{\kappa}
^{2}},
\end{eqnarray}}
where
{
\begin{eqnarray}
\label{eq39} {\cal{M}}=6{H}^{2}+4H\,H'-{\alpha}\,{\kappa}^ {2}\,,\quad
{\cal{X}}=\sqrt {1-{\frac {{\alpha}{\kappa}^{2}\,{\cal{M}} }{ \big(
3{H}^{2}+2\,H\,H' \big) ^{2}}}}\,.
\end{eqnarray}}
The slow-roll parameters in terms of the Hubble parameter and its
derivatives are given by the following expressions
{
\begin{eqnarray}
\label{eq40}
\epsilon=\frac{3}{2}-\frac{3}{2}\,\frac {1}{{\kappa}^{2}} \left( {\frac {{\cal{X}}\,{{\cal{N}}}^{2}}{{\cal{M}}}}+{
		\frac {  {\cal{N}}
			^{2}}{{\cal{M}}}} \right) \Bigg( {\frac {{{\cal{N}}}^{2}}{{\kappa}^{2}{\cal{M}}{\cal{X}}}
	}+{\frac {{{\cal{N}}}^{2}}{{\kappa}^{2}{\cal{M}}}}\hspace{3cm}\nonumber\\+{\frac {9\,  H ^{4}{\cal{X}}-3\,  H  ^{2}
			{\cal{X}}\,{\kappa}^{2}\alpha-9\,  H ^{4}-
			12  H  ^{3}H'
			 -4\,  H ^{2} 
		H'^{2}{\cal{X}}-4\,
			 H  ^{2}  H'^{2}}{{\kappa}^{2}{\cal{M}}{\cal{X}}}} \Bigg) ^{-1}
\end{eqnarray}}
where $H\equiv H(N)$ and {${\cal{N}}=2HH'+3H^2$}
{
\begin{eqnarray}
\label{eq41}\eta=-3\,{\frac {{\cal{N}}  \Big[ {\cal{N}}  \left( \frac{1}{2}\, \left( {\cal{X}} +1 \right) ^{2}
	{\cal{N}} ^{2}+{\cal{K}} 
		\left(  {\cal{X}}  +\frac{1}{2} \right)  \right)  {\cal{X}}'  + \left(  {\cal{X}}  +1
		\right)  \left( {\cal{K}}  {\cal{N}}' -\frac{1}{2}\,{\cal{N}}  {\cal{K}}'  \right) {\cal{X}} 
		\Big] }{\epsilon\,H\, \Big[  \left( {\cal{X}}  +1 \right)  {\cal{N}}  ^{2}+{\cal{K}}   \Big] 
		^{2}}}
\,,
\end{eqnarray}}

with
{
\begin{eqnarray}
\label{eq41b}{\cal{K}}=9  H ^{4}{\cal{X}} -3\,
 H ^{2}{\cal{X}} {\kappa}
^{2}\alpha-9\,  H ^{4}-12\, H
^{3}H'  -4\,  H ^{2}  H'^{2}{\cal{X}}  -4\,  H ^{2} H'^{2}
\,,
\end{eqnarray}}

{
\begin{eqnarray}
\label{eq42}s= \frac{1}{4}\frac { \left[  \bigg( -\frac{\lambda}{2} \left( 
\alpha+2V \right) V' +V \lambda'  \left( \alpha-V 
		\right)  \bigg) \sqrt {V-\alpha }+\frac{V^{\frac{3}{2}}}{2}  V' \alpha \right] {\alpha}^{2}}{\sqrt {V
 } \left( \lambda \sqrt {V-\alpha }-\frac{\alpha}{2}\sqrt {V  } \right) 
		\left( \alpha-V  \right)  \left( \lambda  \left( \alpha-V  \right) \sqrt {V-\alpha}+\frac{\alpha}{2} V^{\frac{3}{2}} \right) }
\,.
\end{eqnarray}}
In the intermediate inflation, the slow-roll parameter $\epsilon$ takes the following form
{
\begin{eqnarray}
\label{eq43}\epsilon=\frac{1}{12}\frac {1}{{N}^{2}\beta} \Bigg[  \Bigg( 2\,\alpha\,{b}^{-2\,{
	\beta}^{-1}} \left( \beta-1 \right) {N}^{{\frac {2+2\,\beta}{\beta}}}-
12\, \bigg( -\frac{1}{4}{b}^{-2\,{\beta}^{-1}}{N}^{{\frac {3\,\beta+2}{\beta
}}}\alpha+ \left( -\frac{2}{3}+ \left( N+\frac{2}{3} \right) \beta \right)\nonumber\\  \left( 
	\beta-1 \right) {N}^{3} \bigg) \beta \Bigg) {{\rm e}^{{\frac {-2\,
					\ln  \left( N \right) +2\,\ln  \left( b \right) }{\beta}}}}-3\,{N}^{3}
	\alpha\,\beta \Bigg] \nonumber\\ \Bigg[  \left( -\frac{2}{3}+ \left( N+\frac{2}{3} \right) \beta
	\right) \beta\,{b}^{2}{{\rm e}^{2\,{\frac { \left( \beta-1 \right) 
					\left( \ln  \left( N \right) -\ln  \left( b \right)  \right) }{\beta}
	}}}-\frac{1}{6}\,N\alpha \Bigg] ^{-1}
\,.
\end{eqnarray}}

\begin{figure}[]
\begin{center}
\includegraphics[scale=0.5]{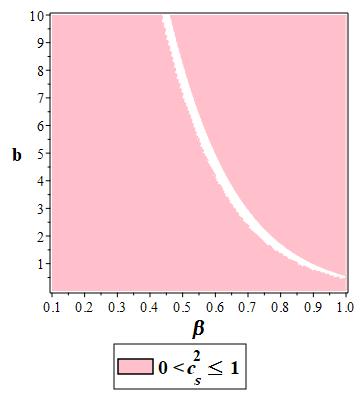}
\includegraphics[scale=0.5]{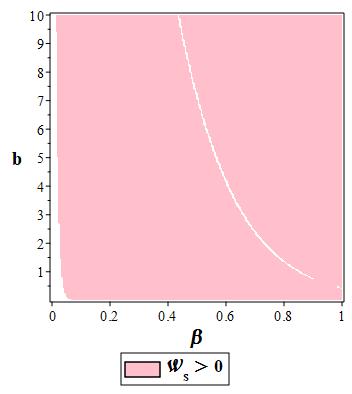}
\end{center}
\caption{\small {Ranges of the model's parameters that
lead to the gradient instability-free intermediate MDBI inflation
(left panel) and the ghost instability-free intermediate MDBI model
(right panel). {These figures have been plotted for $N=60$.}}}
\label{fig1}
\end{figure}

\begin{figure}[]
\begin{center}
\includegraphics[scale=0.5]{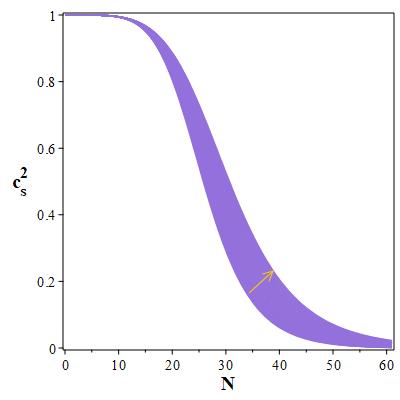}
\includegraphics[scale=0.5]{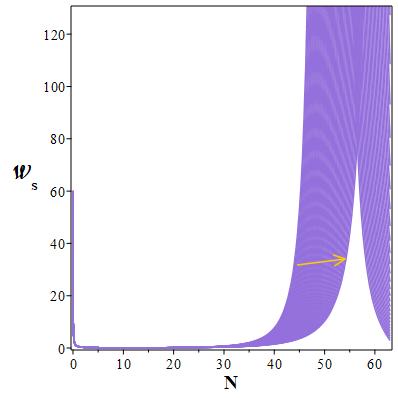}
\end{center}
\caption{\small {The evolution of $c_{s}^{2}$ and ${\cal{W}}_{s}$ of the intermediate MDBI model versus the e-folds number during inflation, with $b=10$ and $0.340<\beta<0.400$. The arrow shows the direction in which the parameter $\beta$ increases.}}
\label{fig2}
\end{figure}

\begin{figure}[]
\begin{center}
\includegraphics[scale=0.5]{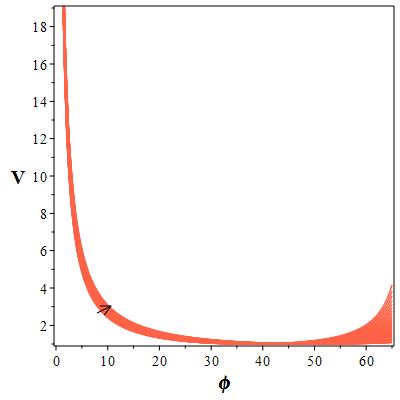}
\end{center}
\caption{\small {The evolution of the potential in the intermediate MDBI model versus the scalar filed $\phi$, with $b=10$ and $0.340<\beta<0.400$. The arrow shows the direction in which the parameter $\beta$ increases. }}
\label{fig3}
\end{figure}
{
The slow-roll parameters $\eta$ and $s$ are very long and complicated, so
we avoid writing these parameters here.} By using the above equations,
we can express the perturbation parameters in terms of the model's
parameters. In Ref.~\cite{Noz19}, we have shown that the MDBI model
with power-law scale factor is free of ghost and gradient
instabilities. Now, in this paper, we show numerically that the
intermediate MDBI model also, in some ranges of the model's
parameter space, is free of the instabilities. In this regard, by
using equations (\ref{eq37}) and (\ref{eq38}) and by assuming
${\cal{F}}=V^{-1}$, we can obtain the sound speed (equation
(\ref{eq10})) in the intermediate MDBI model. Any range of the
parameter space in which we have $c_{s}^{2}>0$, leads to the
gradient instability-free intermediate MDBI model. Note that,
another constraint on the sound speed is $c_{s}\leq c$, where $c$ is
the value of the local speed of light. This constraint is required
from causality. If we consider the scale factor (\ref{eq35}) and
perform some numerical analysis, we find that the range of the
parameter space lead to $0<c_{s}^{2}\leq 1$ (with $c\equiv 1$). The
result is shown in figure 1. {Note that, in this figure and
forthcoming figures we have adopted $\kappa=1$ and $\alpha=1$}. The pink region in the left panel of figure 1 shows the
ranges of the parameters $\beta$ and $b$ leading to the gradient
instability-free intermediate MDBI model. Now, we study
${\cal{W}}_{s}$ to see if the intermediate MDBI model is free of
ghost instability. To this end, we use equations (\ref{eq18}),
(\ref{eq37}), (\ref{eq38}) and (\ref{eq39}). By performing the
numerical analysis, we find the result shown in the right panel of
figure 1. In summary, the intermediate MDBI model in some ranges of
its parameter space is free of gradient and ghost instabilities,
which is a good result.

To ensure there are no instabilities during the whole inflationary
evolution, we plot parameters $c_{s}^{2}$ and ${\cal{W}}_{s}$ versus
the e-folds number $N$, for some sample values of the model's
parameter. The results are shown in figure 2. This figure has been
plotted with $b=10$ and $0.340<\beta<0.400$ (we see in the next section that, these adopted values of $b$ and $\beta$ are observationally viable). As this figure shows,
both conditions $0<c_{s}^{2}\leq 1$ and ${\cal{W}}_{s}>0$ are
satisfied during inflation.

Now, we study the behavior of the potential (\ref{eq37}) versus the
scalar field. To this end, we consider $N=\int Hdt=\int Hd\phi$
(where we have used the constraint (\ref{eq1})) and from equations
(\ref{eq36}) and (\ref{eq37}) we obtain
{
\begin{eqnarray}
\label{eq45} V(\phi)= \frac{1}{6}\,{\frac
{{\phi}^{3\,\beta-2}{\beta}^{2}{b}^{2} \left( 3\,\beta\,b\,{
\phi}^{\beta}+2\,\beta-2 \right) ^{2}}{{\kappa}^{2} \left(
\frac{2}{3}\,\beta \,b \left( \beta-1 \right)
{\phi}^{2\,\beta}+{\phi}^{3\,\beta}{\beta}^
{2}{b}^{2}-\frac{\alpha}{6}\,{\kappa}^{2}{\phi}^{\beta+2} \right) }}\,.
\end{eqnarray}}
By using this equation, we can perform a numerical study on the
evolution of the potential versus the scalar field. The result is
shown in figure 3. At the initial times, where
inflation happens, the potential is large, and also the friction term
$3H\dot{\phi}$.

\begin{figure}[]
\begin{center}
\includegraphics[scale=0.5]{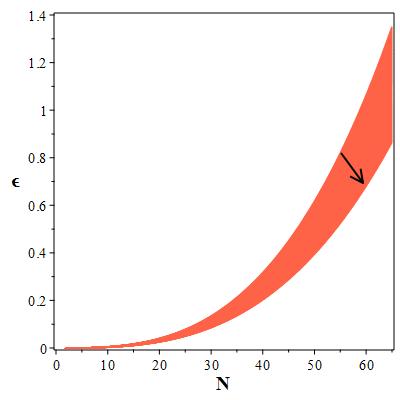}
\end{center}
\caption{\small {The evolution of the slow-roll parameter $\epsilon$ of the intermediate MDBI model versus the e-folds number during inflation, with $b=10$ and $0.340<\beta<0.400$. The arrow shows the direction in which the parameter $\beta$ increases.}}
\label{fig4}
\end{figure}

It is worth checking if inflation ends in this setup. The inflation
ends when one of the slow-roll parameters reaches unity. In this
regard, we study the first slow-roll parameter $\epsilon$ versus the
e-folds number $N$. The result is shown in figure 4. As the figure
shows, in this model inflation ends after about $60$ e-folds (which
is also corresponding to the minimum of the potential) and the
graceful exit of inflation towards the matter-dominated era can be
achieved. Note that, since the scale factor in the intermediate
inflation evolves faster than the one in the power-law inflation and
slower than the one in the exponential inflation, the same situation
happens for the slow-roll parameter $\epsilon$. In this model also,
it is possible to have the seeds for the observed dark matter. In fact, by considering the conservation of the
energy-momentum tensor (as $\nabla^{\mu}\,T_{\mu\nu}=0$), we obtain
\begin{eqnarray}
\label{eq46}
\frac{1}{\sqrt{-g}}\partial_{\mu}\left[\sqrt{-g}\left(-\frac{\partial^{\mu}\phi}
{\sqrt{1+{\alpha}{\cal{F}}\partial^{\mu}\phi\,\partial_{\mu}\phi}}+2\lambda
g^{\mu\nu}\partial_{\nu}\phi\right)\right]=-V'+\frac{{\cal{F}}'}{{\cal{F}}^{2}}\left[\frac{2+3{\alpha}{\cal{F}}g^{\mu\nu}\partial_{\mu}\phi\,
\partial_{\nu}\phi}{2\sqrt{1+{\alpha}{\cal{F}}g^{\mu\nu}\partial_{\mu}\phi\,
\partial_{\nu}\phi}}\right]\,.
\end{eqnarray}
By considering the constraint (\ref{eq2}) and
$H^{2}=\frac{\kappa^{2}}{3}\rho$, form equations (\ref{eq6}) and
(\ref{eq46}) we have
\begin{eqnarray}
\label{eq47}
\frac{1}{\sqrt{-g}}\partial_{\mu}\left[\sqrt{-g}\left(\rho-V-{\cal{F}}^{-1}\sqrt{1-{\alpha}{\cal{F}}}\right)\right]=
-V'+\frac{{\cal{F}}'}{{\cal{F}}^{2}}\left[\frac{2-3{\alpha}{\cal{F}}}{2\sqrt{1-{\alpha}{\cal{F}}}}\right]\,.
\end{eqnarray}

After the end of the inflation, at the moment we reach the minimum
of the potential, we have $V={\cal{F}}^{-1}=constant$. At that
point, the slope of the potential is zero and so is the right-hand side
of the equation (\ref{eq46}). In this case, we have
\begin{eqnarray}
\label{eq48}
\frac{1}{\sqrt{-g}}\partial_{\mu}\left[\sqrt{-g}\left(\rho-V-{\cal{F}}^{-1}\sqrt{1-{\alpha}{\cal{F}}}\right)\right]=
0\,,
\end{eqnarray}
and therefore
{
\begin{eqnarray}
\label{eq49} \rho=
\frac{{\cal{C}}}{a^{3}}+V+{\cal{F}}^{-1}\sqrt{1-\alpha{\cal{F}}}\,,
\end{eqnarray}}
{where ${\cal{C}}$ is a constant with dimension $[M]^{4}$}. In this way, the seeds of the dark
matter are obtained and the term $\frac{{\cal{C}}}{a^{3}}$ determines
the amount of dark matter in the intermediate MDBI model. Note
that, although in the inflation era (where the right-hand side of
the equation (\ref{eq47}) is not zero) it is possible to have a term
like $\frac{{\cal{C}}}{a^{3}}$, this term is diluted away quickly.
However, after the end of inflation, this term has an important role. Also, by considering the gravitational particle production at the end of inflation, it is possible to get the observed radiation and baryons in the universe~\cite{For87}. Although the potential is a function of time and increases again, the gravitational created particles dominate the potential term near the minimum. Then, by increasing the time, the potential term (and also, ${\cal{F}}^{-1}\sqrt{1-{\cal{F}}}$ term) dominates the particles and probably becomes the dark energy component leading to the late-time acceleration of the universe.

In the next section, we study the observational viability of this
model with the Planck2018 data.

\section{Comparing with the Planck2018 Observational Data}
When an inflation model is constructed, it is important to check if
its results are consistent with observational data. The
observational data give some constraints on the perturbation
parameters such as the scalar spectral index, the tensor spectral
index, and the tensor-to-scalar ratio. Also, the observational data
sets constraints on the amplitudes of the non-gaussianity in the
equilateral configuration. Therefore, by studying
these parameters in an inflation model and comparing the results
with observational data, we can explore the viability of the model.
The constraint on the scalar spectral index, from Planck2018 TT, TE,
EE+lowE+lensing +BAO +BK14 data, based on
$\Lambda$CDM$+r+\frac{dn_{s}}{d\ln k}$ model, is $n_{s}=0.9658\pm
0.0038$. This dataset gives the constraint on the tensor-to-scalar
ratio as $r<0.072$. Also, Planck2018 TT, TE, EE
+lowE+lensing+BK14+BAO+LIGO and Virgo2016 constraint on the tensor
spectral index is as $-0.62<n_{T}<0.53$. By using these data, we can
obtain some constraints on the intermediate parameters $\beta$ and
$b$. By substituting equations (\ref{eq40}), (\ref{eq41}) and
(\ref{eq42}) for the intermediate inflation in the equations
(\ref{eq19}), (\ref{eq22}), (\ref{eq23}), we obtain the
tensor-to-scalar ratio, the scalar spectral index and the tensor
spectral index in terms of the parameters $\beta$ and $b$. Then, we
perform numerical analysis on these parameters and compare the
results with observational data. Our numerical analysis shows that,
for $0<b\leq10$, depending on the values of $b$, the scalar spectral
index in the intermediate MDBI model is consistent with
observational data if $0.345< \beta< 0.387$. This is shown in the
left-upper panel of figure 5. The tensor-to-scalar ratio in this
model is consistent with observational data if, depending on the
values of $b$, $0.341< \beta< 1$. This is shown in the right-upper
panel of figure 5. Also, we have found that the range $0.044< \beta<
1$ leads to the observationally viable values of the tensor spectral
index in the intermediate MDBI model. This is shown in the lower
panel of figure 5. We have also studied the tensor-to-scalar ratio
versus the scalar spectral index in the background of Planck2018 TT,
TE, EE+lowE+lensing +BAO +BK14 data at $68\%$ CL and $95\%$ CL. The
results are shown in figure 6. From this figure, we have
obtained some constraints on the parameter $\beta$ that are
summarized in table 1. The tensor-to-scalar ratio versus the tensor
spectral index in the background of Planck2018 TT, TE,
EE+lowE+lensing +BAO +BK14 data (figure 7) is another studied case
that gives some more constraints presented in table 1.

\begin{figure}[]
\begin{center}
\includegraphics[scale=0.5]{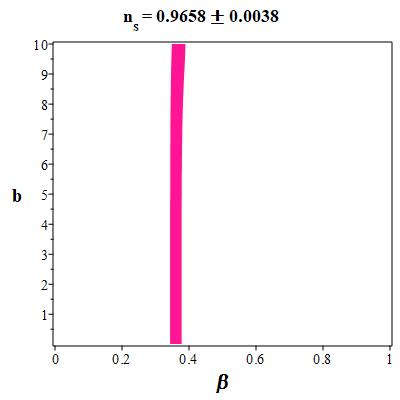}
\includegraphics[scale=0.5]{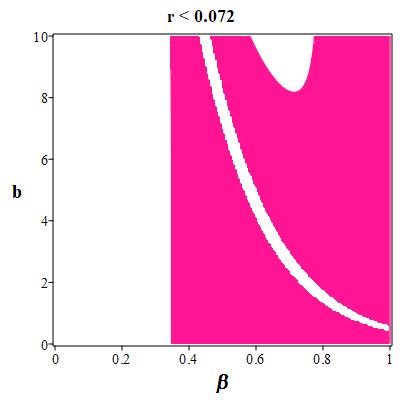}
\includegraphics[scale=0.5]{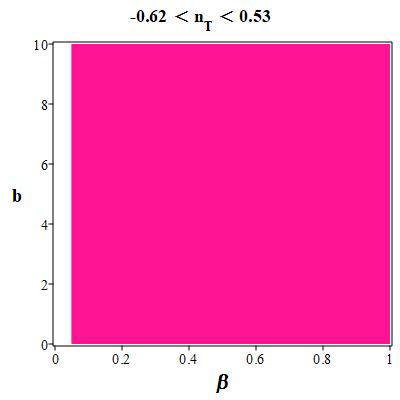}
\end{center}
\caption{\small {The upper panels show the ranges of the
model's parameter space that lead to the observationally viable
values of the scalar spectral index (the left one) and tensor-to-scalar
ratio (the right one), obtained from Planck2018 TT, TE, EE+lowE+lensing
+BAO +BK14 data. The lower panel shows the range of the model's
parameter space that leads to the observationally viable values of
the tensor spectral index, obtained from Planck2018 TT, TE, EE
+lowE+lensing+BK14+BAO+LIGO and Virgo2016 data. {These figures have been plotted for $N=60$.}}}
\label{fig5}
\end{figure}

\begin{figure}[]
\begin{center}
\includegraphics[scale=0.5]{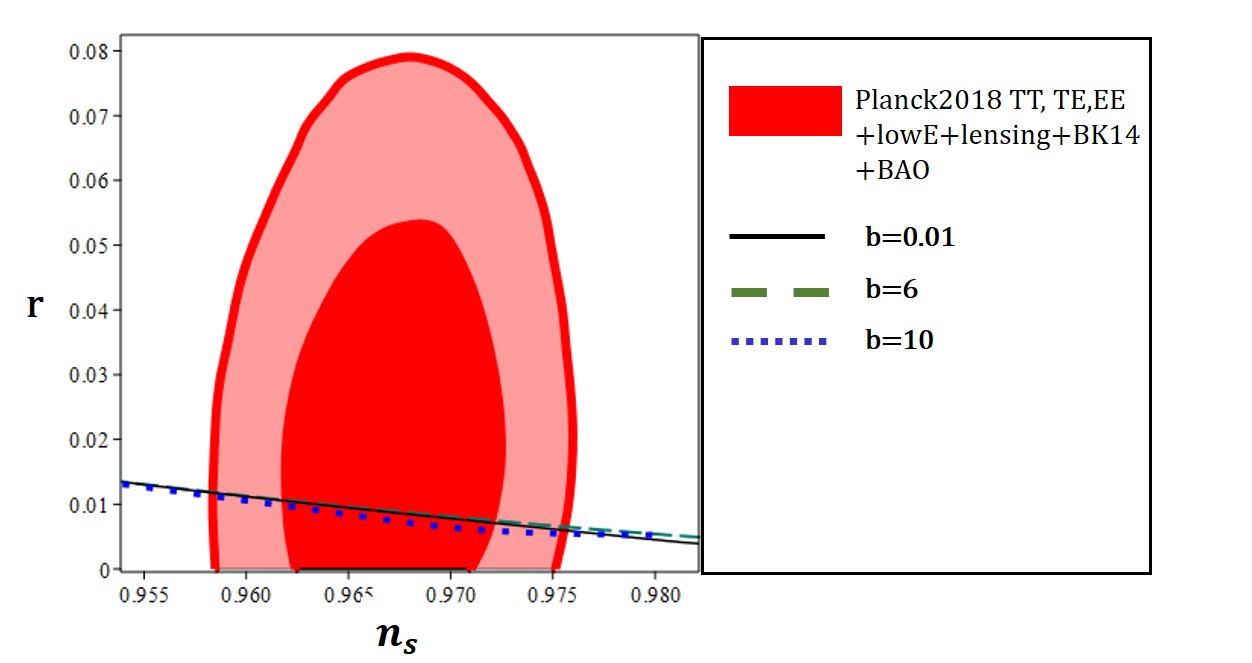}
\end{center}
\caption{\small {Tensor-to-scalar ratio versus the
scalar spectral index for the intermediate MDBI model. {To plot this figure, we have adopted $0<\beta<1$ and $N=60$.}}}
\label{fig6}
\end{figure}

\begin{figure}[]
\begin{center}
\includegraphics[scale=0.5]{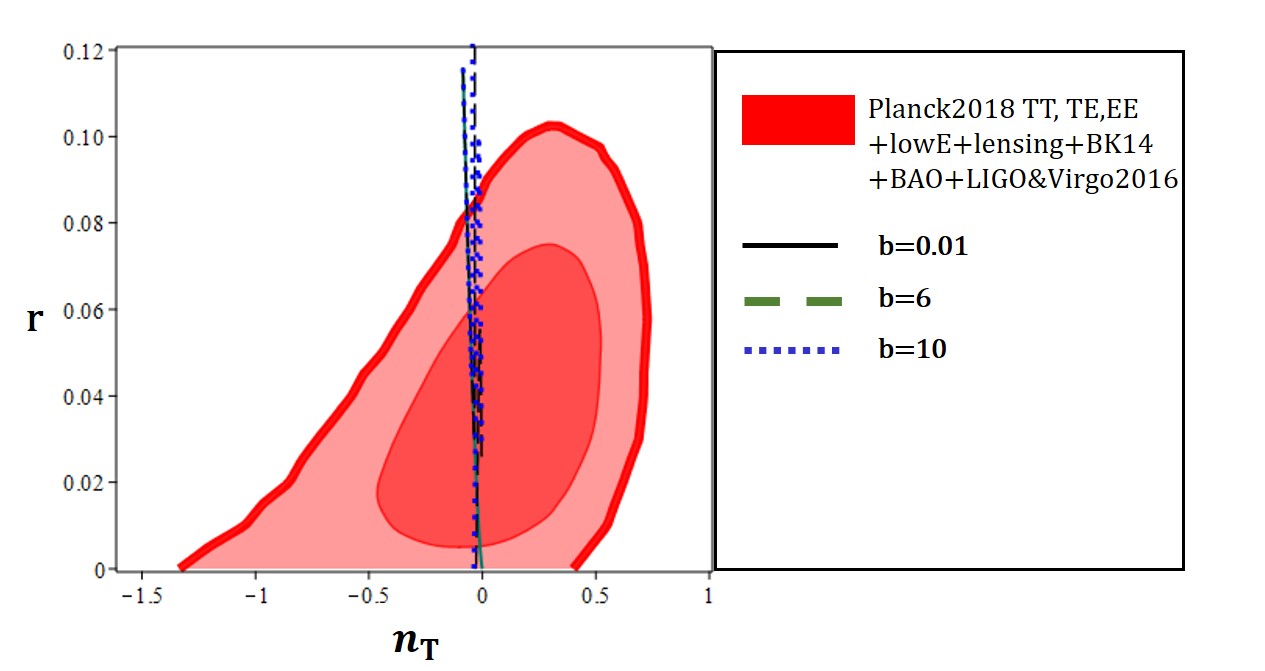}
\end{center}
\caption{\small {Tensor-to-scalar ratio versus the
tensor spectral index for the intermediate MDBI model. {To plot this figure, we have adopted $0<\beta<1$ and $N=60$.}}}
\label{fig7}
\end{figure}

\begin{table*}
\tiny \caption{\small{\label{tab:1} The ranges of the parameter
$\beta$ in which the tensor-to-scalar ratio, the scalar spectral
index, and the tensor spectral index of the intermediate MDBI model
are consistent with different data sets.}}
\begin{center}
\begin{tabular}{cccccc}
\\ \hline \hline \\ & Planck2018 TT,TE,EE+lowE & Planck2018 TT,TE,EE+lowE&Planck2018 TT,TE,EE+lowE&Planck2018 TT,TE,EE+lowE
\\
& +lensing+BK14+BAO &
+lensing+BK14+BAO&lensing+BK14+BAO&lensing+BK14+BAO
\\
&  & &+LIGO$\&$Virgo2016 &LIGO$\&$Virgo2016
\\
\hline \\$b$& $68\%$ CL & $95\%$ CL &$68\%$ CL & $95\%$ CL
\\
\hline\hline \\  $0.01$& $0.345< \beta<0.387 $&$0.334<
\beta<0.403$&$0.375< \beta\leq 0.771$ & $0.329< \beta$\\ \\
\hline
\\$6$&$0.346<\beta<0.390$&$0.333<\beta <0.410$ &$0.375< \beta<0.499 $&$0.327< \beta <0.500$
\\ \\
&&&$0.557< \beta <0.980 $& $0.499<\beta$\\ \\ \hline\\
$10$&$0.350< \beta <0.396 $&$0.337< \beta <0.404$&$0.369< \beta
<0.429 $  &$0.327< \beta
<0.431 $\\ \\
&&&$0.477< \beta <0.567 $&$0.431<\beta <0.605 $ \\ \\
&&&$0.799< \beta <0.981 $& $0.730<\beta$\\ \\
\hline \hline
\end{tabular}
\end{center}
\end{table*}

Another important property in studying an inflation model is the
non-gaussian feature of the primordial perturbations. As we
mentioned earlier, the amplitude of the non-gaussianity is related
to the sound speed of the perturbations and in this way, it is
related to the model's parameter space. In this regard, studying the non-gaussian features of the perturbations in the inflation
models helps us get more information about the models. We study the
amplitude of the non-gaussianity in equilateral configurations. By
considering the constraints on the model's parameters obtained
from the observational viability of the scalar and tensor spectral
index and the tensor-to-scalar ratio from Planck2018, we analyze the
non-gaussianity in the intermediate MDBI inflation numerically.
Here, we consider the non-gaussianity parameter $f^{equil}$, which is
related to the sound speed by equation (\ref{eq34}). By using
equations (\ref{eq10}), (\ref{eq37}) and (\ref{eq38}) and also
assuming ${\cal{F}}^{-1}=V$, we relate the square of the sound speed
of the primordial perturbations to the model's parameters. The
square of the sound speed is also related to the tensor-to-scalar
ratio via equation (\ref{eq23}). In this regard, the constraints on
$r$ set some constraints on $c_{s}^{2}$ that is shown in figure
8.

\begin{figure}[]
\begin{center}
\includegraphics[scale=0.5]{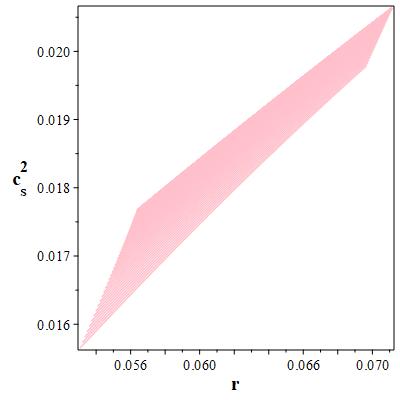}
\end{center}
\caption{\small {The square of the sound speed versus
the tensor-to-scalar ratio for the intermediate MDBI model. {To plot this figure, we have adopted $0.340<\beta<0.4$, $0<b<10$ and $N=60$.}}}
\label{fig8}
\end{figure}

\begin{figure}[]
	\begin{center}
		\includegraphics[scale=0.5]{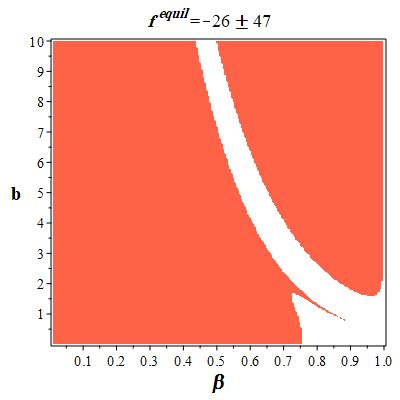}
		\includegraphics[scale=0.5]{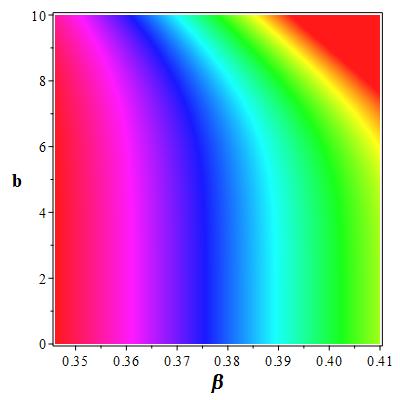}
		\includegraphics[scale=0.5]{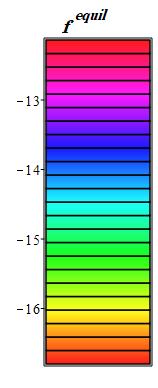}
	\end{center}
	\caption{\small {The left panel shows the ranges of the intermediate MDBI model's parameters leading to the observationally viable values of the equilateral non-gaussianity. The right panels show the prediction of the intermediate MDBI model for equilateral non-gaussianity. {These figures have been plotted for $N=60$.}}}
	\label{fig5}
\end{figure}

The planck2018 combined temperature and polarization
analysis gives the constraint on the sound speed of the DBI model as
$c_{s}^{DBI}\geq 0.086$, at $95\%$ CL. Also, this data gives the
constraint on the sound speed of the general $P(X,\phi)$ model
(where, $X=-\frac{1}{2}\partial_{\nu}\phi\,\partial^{\nu}\phi$) as
$c_{s}\geq 0.021$, at $95\%$ CL. According to our numerical
analysis, shown in figure 8, the values of the sound speed in
our intermediate MDBI model are consistent with both constraints.

By having the constraints on the model's parameters $\beta$, $b$ and
therefore $c_{s}^{2}$, we can predict the non-gaussian feature of
the primordial perturbations in the intermediate MDBI model from
equation (\ref{eq34}).
The Planck2018 data gives the constraint on the equilateral
configuration of the non-gaussianity as $f^{equil}=-26\pm 47$. By
using this constraint, we can find the range of the parameter space
leading to the observationally viable values of the equilateral
non-gaussianity in the intermediate MDBI model. The result is shown
in the left panel of figure 9.
Also, the right panels of figure 9 show the prediction of the
model for the equilateral amplitude of the non-gaussianity, where we
have used the observationally viable ranges of the parameters $b$
and $\beta$ which are obtained from the comparing of the
tensor-to-scalar ratio and scalar spectral index with Planck2018 TT,
TE, EE+lowE+lensing +BAO +BK14 data at $95\%$ CL.{
Note that, we have used the same ranges of the parameters to plot figures 6 and 7. However, plot 6 is $r-n_s$ behavior and plot 7 is $r-n_T$ behavior. From the slope of the $r-n_s$ plot, we see that it is possible to have larger values of r, but these larger values are corresponding to smaller values of $n_s$, which are not observationally viable. Therefore, we haven’t shown those parts of the plot which are not consistent with observational data. However, the situation is different for $r-n_T$. As figure 7 shows, if we consider the larger values of r, the parameter $n_T$ is still consistent with observational data.}

\begin{table*}
\small \caption{\small{\label{tab:2} The prediction of the model for
the equilateral amplitude of the non-gaussianity, obtained from the
observationally viable values of the parameter $\beta$ at $68\%$ CL
and $95\%$ CL.}}
\begin{center}
\begin{tabular}{cccccccccccc}
\\ \hline \hline \\ & $c_{s}^{2}$ &&  $c_{s}^{2}$ && $f_{NL}^{equil}$ & &$f_{NL}^{equil}$
\\
\hline \\$b$& $68\%$ CL && $95\%$ CL &&$68\%$ CL && $95\%$ CL
\\
\hline\hline \\  $0.01$& $[0.017,0.021] $&&$[0.016,0.022]$&&$[-15.1,-12.2]$ && $[-16.1,-11.6]$\\ \\
\hline
\\$6$&$[0.016,0.020]$&&$[0.015,0.022]$ &&$[-16.1,-12.8]
$&&$[-17.2,-11.6]$
\\ \\ \hline\\
$10$&$[0.014,0.019] $&&$[0.013,0.021]$&&$[-18.4,-13.5]$
&&$[-19.9,-12.2] $\\ \\
\hline \hline
\end{tabular}
\end{center}
\end{table*}

Up to here, we have numerically studied the perturbation parameters
$n_{s}$, $n_{T}$ and $r$ and also the non-gaussianity parameter
$f_{NL}^{equil}$, to find some constraints on the model's
parameters. According to the obtained results, for every parameter,
there are some ranges in the parameter space which make the model
observationally viable. However, we are interested in the case where
all parameters are consistent with Planck2018 data in the same range.
By this, we mean that it is interesting to find a range for the
parameters $\beta$ and $b$ where we have an instabilities-free and
observationally viable intermediate MDB model. Our data analysis to
obtain such ranges shows that by $0<b\leq10$ and
$0.345<\beta<0.387$, it is possible to have an instabilities-free
intermediate MDBI model that gives the observationally viable
perturbations. Also, with these ranges, the values of the
equilateral amplitude of the non-gaussianity are consistent with
observational data.

\section{A Short Discussion on the Difference between the Power-Law MDBI and Intermediate MDBI Models}
In Ref.~\cite{Noz19}, we have studied the power-law DBI and power-law MDBI models, where $a=a_{0}t^{n}$, with details. We have obtained the main inflation and perturbation parameters in both models and performed a numerical
analysis on those parameters. According to our analysis in Ref.~\cite{Noz19}, the result of the numerical analysis on the perturbation parameters of the power-law DBI model is not consistent with Planck2018 observational data. Then, we have considered the power-law MDBI model. We have shown that the power-law MDBI
model is an instabilities-free model. We have also shown that the
scalar spectral index and the tensor-to-scalar ratio in the
power-law MDBI model are consistent with Planck2018 TT, TE, EE+lowE
+lensing data at $95\%$ CL. However, these perturbation parameters
in the power-law MDBI model are not consistent with Planck2018 TT,
TE, EE+lowE+lensing+BK14+BAO data, where the combination of the
BICEP2/Keck Array 2014 and Planck2018 data is considered. Note that,
in the case of constant sound speed, it is possible to get
observational consistency, however, our attention is on the varying
sound speed case. In this paper, We have shown that if we consider an intermediate MDBI model, the perturbation parameters $n_{s}$ and
$r$ are consistent with Planck2018 TT, TE, EE+lowE+lensing+BK14+BAO
data at both $68\%$ CL and $95\%$ CL. This is an interesting
advantage of the intermediate MDBI model over the power-law MDBI
model.

We have also explored the non-gaussian feature of the primordial
perturbation in the power-law DBI and power-law MDBI models, in Ref.~\cite{Noz19}. In that paper, it has been shown that the amplitude of the primordial non-gaussianity in the power-law DBI model is too large to be consistent with Planck2018 observational data. Also,
we have shown that the prediction of the power-law MDBI model for
the amplitude of the equilateral non-gaussianity is very small (of
the order of $10^{-4}$). However, as we have seen in the current
paper, the equilateral non-gaussianity in the intermediate MBI model
is in the range $-16.7<f^{equil}<-12.5$. This is a result that is
consistent with planck2018 data.

It seems that, by considering both perturbation and non-gaussianity
parameters, the intermediate MDBI model is consistent with
Planck2018 data and therefore is more favorable.

\section{Summary and Conclusion}

Recently, it has been shown that to have a ghost and gradient
instabilities-free mimetic gravity model, one can consider a DBI-like term in the action of the mimetic gravity and adopt a
power-law scale factor. In this paper, we have considered a MDBI
model with intermediate scale factor as $a=a_{0}\exp(bt^{\beta})$.
In this regard, we have studied the intermediate inflation in the
MDBI model. We have shown that, with the intermediate MDBI model, it
is possible to have a mimetic gravity model that is free of ghost
and gradient instabilities in some ranges of the intermediate
parameters $b$ and $\beta$. This means that, in those ranges of the
parameters, we have $0<c_{s}^{2}\leq 1$ and ${\cal{W}}_{s}>0$. To
seek the observational viability of the models, we have studied
the perturbation and non-gaussianity parameters of this model and
compare the results with observational data. From Planck2018 TT, TE,
EE+lowE+lensing +BAO +BK14 data, we have the value of the scalar
spectral index as $n_{s}=0.9658\pm0.0038$. This implies that, for
$0<b\leq10$, the constraint on the $\beta$ is as $0.345<
\beta<0.387$. Planck2018 TT, TE, EE+lowE+lensing +BAO +BK14
constraint on the tensor-to-scalar ratio is $r<0.072$, leading to
$0.341< \beta< 1$ for $0<b\leq10$. From Planck2018 TT, TE, EE
+lowE+lensing+BK14+BAO+LIGO and Virgo2016 data, the constraint on
the tensor spectral index is $-0.62<n_{T}<0.53$. In this regard, we
find that for $0<b\leq10$, the range $0.044< \beta< 1$ leads to the
observationally viable values of the tensor spectral index in the
intermediate MDBI model. We have also studied $r-n_{s}$ and
$r-n_{T}$ behaviors of the intermediate MDBI model in comparison to
the observational data at $68\%$ CL and $95\%$ CL and found some
constraints summarized in the tables.

Another important aspect of the inflation models is the non-gaussian
feature of the primordial perturbations. In this paper, we have
studied the non-gaussianity in the equilateral configuration. We
have considered the $k_{1}=k_{2}=k_{3}$ limit, where the equilateral
configuration has a peak. By studying this configuration of the
non-gaussianity and considering the observationally viable ranges of
the parameters $b$ and $\beta$, we have predicted the amplitudes of
the non-gaussianity in our intermediate MDBI model.

As a summary, we have shown that our proposed intermediate MDBI
model with $0<b\leq10$ and $0.345<\beta<0.387$, is
instabilities-free and gives the observationally viable perturbation
and non-gaussianity parameters.

{
{\bf Acknowledgement}\\
We thank the referee for the very insightful comments that have
improved the quality of the paper considerably.}\\

\end{document}